%% file: main.tex
\documentclass[a4paper,11pt]{amsart}

\usepackage{graphicx}
\usepackage{a4wide,verbatim}
\usepackage[dvipsnames]{xcolor}
\usepackage{epstopdf, epsfig}
\usepackage{amsaddr}
\usepackage{amsaddr,amsmath,amsfonts,amssymb,amsgen,amsbsy,dsfont}
\usepackage{mathtools}
\usepackage{bm}
\usepackage{floatrow}
\usepackage{subcaption}
\usepackage{caption}
\usepackage{hyperref}
\usepackage{upgreek}
\usepackage{tikz}
\usepackage{dsfont}
\usepackage{csvsimple}
\usepackage{multirow}
\usepackage{array}
\usepackage{mathabx}
\usepackage{adjustbox}

\newcommand{\depth}{d}                              
\newcommand{\sur}[1]{{#1}_\mathrm{s}}                    
\renewcommand{\bot}[1]{{#1}_\mathrm{b}}                  

\newcommand{\rf}[1]{(\ref{#1})}
\newcommand{\eqdef}{\vcentcolon=}

\DeclareMathOperator*{\argmin}{arg\,min}
\newcommand{\obs}[1]{{#1}_\mathrm{b}^\mathrm{obs}}                  

\newcommand{\ud}{\mathrm{d}}

\newcommand{\ue}{\mathrm{e}}
\newcommand{\ui}{\mathrm{i}}
\newcommand{\upi}{{\pi}}

\newcommand{\scal}{\boldsymbol{\cdot}}              

\newcommand{\half}{{\textstyle{\frac{1}{2}}}}

\renewcommand{\Re}{\operatorname{Re}}
\renewcommand{\Im}{\operatorname{Im}}

\newlength{\intwidth}

\newcommand{\plog}[2]{\mbox{L{\scriptsize{}}}_{#1}#2}    

\begin{document}\sloppy

\title[Optimal reconstruction of water-waves from pressure measurements]{Optimal reconstruction of water-waves from noisy pressure measurements at the seabed}



\author[J. Labarbe]{Joris Labarbe}
\address{Universit\'e C\^ote d'Azur, CNRS UMR 7351,  Laboratoire J. A. Dieudonn\'e, 
Parc Valrose, 06108 Nice cedex 2, France}
\email{joris.LABARBE@univ-cotedazur.fr}

\author[A. Vieira]{Alexandre Vieira}
\address{Universit\'e C\^ote d'Azur, CNRS UMR 7351,  Laboratoire J. A. Dieudonn\'e, 
Parc Valrose, 06108 Nice cedex 2, France}
\email{alexandre.VIEIRA@univ-cotedazur.fr}

\author[D. Clamond]{Didier Clamond}
\address{Universit\'e C\^ote d'Azur, CNRS UMR 7351,  Laboratoire J. A. Dieudonn\'e, 
Parc Valrose, 06108 Nice cedex 2, France}
\email{didier.CLAMOND@univ-cotedazur.fr}

\subjclass[]{}

\begin{abstract}
We consider the problem of recovering the surface wave profile from noisy bottom pressure measurements with (\textit{a priori} unknown) arbitrary pressure at the surface.
Without noise, the direct approach developed in \cite{clamond2023steady} provides an effective way to recover the sea surface.
However, the assumption of analyticity for the measurement renders this method inefficient in the presence of noise.
Therefore, we introduce an optimisation procedure based on the minimisation of a distance between a recovered bottom pressure and its measurement.
Such method proves to be well-designed to handle perturbed signals.
We illustrate the effectiveness of this approach in the recovery of gravity-capillary waves from unfiltered noisy data.
\end{abstract}


\maketitle

\section{Introduction}

Monitoring the surface of the oceans is a timely concern for climatic and environmental considerations, especially in coastal regions where large waves represent a risk when approaching the shoreline.
Currently, an efficient way of monitoring the wave motion is lacking, as most current techniques display a poor accuracy in their measurements.
For this reason, in the second half of the last century, some scientists emitted the idea of sending probes at the seabed to measure the pressure and reconstruct the surface profile from these data \cite{BTT69}.
This procedure allows to determine the shape of water-waves without intrusive observations, although it necessitates to solve an inverse problem within an unknown domain.
Such problem is notoriously challenging, notably because of its ill-posedness (any disturbance grows exponentially from the bottom to the free surface).

Historically, the first attempt to solve this problem relied on hydrostatic theory and assumed the surface elevation as being proportional to the weight of the water column. 
Later, from a linear approximation of the problem, an explicit expression for the surface was given in terms of a certain pressure transfer function \cite{LW85}.
Noticeably, a recent work derived an exact formula for steady rotational waves using linear theory \cite{HT16}.
However, both these assumptions were not suited to describe large waves or of non-trivial profile (i.e. not sinusoidal).
It was only recently that the first results on the full nonlinear problem were obtained \cite{O12}.
This study resulted in a nonlocal integro-differential equation for the surface elevation that was highly difficult to solve in practice without considering asymptotic limits.
Slightly after this work, an efficient procedure for the recovery of nonlinear steady water-waves was formulated \cite{C13,CC13}.
This approach relies on the assumption that bottom pressure can be analytically continued as a holomorphic function, allowing to express the surface elevation as a solution of an algebraic equation.
Numerical investigation of this method was done for extreme waves \cite{CC13} and it was rigorously proved to converge in the case of steady surface gravity waves from a mere fixed-point algorithm \cite{CH20}.
Subsequently, this theory was extended to handle more complex configurations, such as the presence of a linear shear current \cite{CLH23}, the possibility of overhanging waves \cite{LC23} or the influence of arbitrary pressure at the surface \cite{clamond2023steady}.
We note the proof for the existence of nonlinear water-waves with constant vorticity and overturning wave profiles \cite{CSV11,CSV16}.

The method originally developed in \cite{CC13} consistently proved its effectiveness in terms of accuracy, computation time and accessibility, even considering technical difficulties (stagnation points, limiting waves, capillary effects, etc.).
However, the formulation inherently depends on the analyticity of the measurement, which is not the case in real-life observations where the presence of external noise is inevitable. 
Moreover, since we are performing most of the calculations in the complex plane, then it prohibits us to directly extend this approach to the three-dimensional configuration.
For all these reasons, we introduce a new and innovative formulation, based on an optimisation method, to solve this constrained inverse problem with noise in the measurements.

To the best of authors' knowledge, surface wave recovery within an optimisation framework has never been considered in the literature.
As it turns out, one can relate this reconstruction problem to an optimal control approach, where the goal is to find a set of parameters or functions minimising a given cost functional under some dynamical constraints.
(We refer interested readers to \cite{cossu2014introduction} for an introduction on this subject.)
In the current inverse problem, the undetermined function is the surface profile and the cost functional will be the distance to the observation.
In this sense, this problem is close to some shape optimisation or topology optimisation (TO) configurations.
These problems consist in finding the optimal shape of an obstacle immersed in a moving fluid (see \cite{alexandersen2020review,dbouk2017review} and references therein for an overview on the numerical resolution of TO problems applied to several physical settings). 
Although it yields satisfactory results in general, the main drawback of the shape optimisation problem comes from its numerical complexity.
This is because the underlying dynamical constraints are usually discretised using a Finite Element method, which relies on a prescribed mesh adapted to the shape of the obstacle.
In this setting, the shape of the immersed obstacle is updated at each iteration, leading to an expensive recalculation of the mesh \cite{allaire2014shape}.
Thus, all techniques avoiding this costly procedure necessarily enhance the formulation.

The method we consider in this work relies on a predefined discretisation of some analytical expressions we derived in an earlier work, from a boundary integral method \cite{LC23}.
Combining this analytic preprocessing within the optimisation framework has never been explored before and appears to be a suitable and innovative approach for the surface reconstruction problem.
Moreover, the algorithm we develop is easily extendable to a broad range of situations (e.g. adding further physical assumptions or noise in the measurements) and yields an excellent agreement with the sought solution (comparable to the results obtained in \cite{clamond2023steady} using the holomorphic extension).

%


This article is articulated around the presentation, application and comparison of two different methods to reconstruct surface waves when considering noise or not.
Section \ref{secmathdef} introduces the equations of motion and boundary conditions for the system of interest.
Then, sections \ref{secnumapp1} and \ref{secnumapp2} describe in details the direct and optimisation approaches, respectively, and how to implement them numerically in an efficient way.
A numerical example is treated in section \ref{secnumexp} to explore and compare thoroughly these methods.
Finally, we discuss about the results we obtained and on the eventual improvements and future extensions in section \ref{secconclu}.

\section{Problem statement}\label{secmathdef}

We investigate the motion of a steady wave travelling at a constant phase speed $c$ within a fluid layer with a free surface.
The bottom of the layer being flat, we fix ourselves in the Galilean frame of reference moving with the wave, so that motion appears stationary for the observer. 
The fluid is considered inviscid, irrotational and incompressible; we assume the surface and seabed to be impermeable. 
These boundaries are located at $y=\eta(x)$ for the surface and  $y=-\depth$ for the seabed, with $y$ being the vertical coordinate and $\depth$ a constant depth.
We assume the streamwise direction to have an infinite extension.
Therefore, the domain of definition is given by $\Omega = \{(x,y):x\in\mathds{R},-\depth\leqslant y\leqslant\eta(x)\}$. 
When the motion is periodic, we introduce the wave period $L=2\upi/k$ (the limit $k\to0^{+}$ corresponding to a solitary wave) where $k$ is a spatial wavenumber. 
Thus, we define an Eulerian averaging operator to fix the mean water level at $y=0$, i.e.
\begin{equation} 
\label{defmean}
\left< \eta \right> \eqdef \frac{k}{2\upi} \int_{\mathcal{P}} \eta(x) \ud x = 0 ,
\end{equation}
where the path of integration is $\mathcal{P}=[-\upi/k,\upi/k]$.

In the following, we use the subscripts `s' and `b' to denote the restriction of fields at the surface and the bottom, respectively.
Alternatively, the subscript $x$ is used to express horizontal derivatives (e.g. $\eta_x\eqdef\ud\eta/\ud x)$.
We consider the pressure at the surface as an undetermined function of the $x$-coordinate, entering the equations through the dynamical condition (supplemented with kinematic condition).
The velocity field $\bm{u}=(u,v)$ is governed by the steady Euler equations (expressing the conservation of mass and linear momentum) with associated boundary conditions.
It yields
\begin{subequations}\label{eom}
\begin{alignat}{2}
\bm{\nabla} \scal \bm{u} &= 0,   \quad &&\textrm{in} \quad \Omega \\
\bm{u} \scal \bm{\nabla} \bm{u} + \bm{\nabla} p + \bm{g} &= 0,   \quad &&\textrm{in} \quad \Omega \label{euler} \\
\bm{u} \scal \bm{n} - \bm{u} \scal \bm{\nabla} \eta &= 0  \quad &&\textrm{at} \quad y=\eta(x) , \\
p &= \sur{p}(x)  \quad &&\textrm{at} \quad y=\eta(x) , \\
\bm{u} \scal \bm{n} &= 0  \quad &&\textrm{at} \quad y=-d ,
\end{alignat}
\end{subequations}
where $p$ is a relative pressure with zero mean-value at the surface (i.e. $\langle \sur{p}\rangle =0$) and scaled with the constant background density $\rho_0$. 
The unit normal vector $\bm{n}$ is directed outward and $\bm{g}=(0,g)$ represents the restoring gravity force, with $g$ the acceleration due to gravity (acting downwards).
By integrating equation \rf{euler}, we recover the well-known Bernoulli principle (expressing the conservation of mechanical energy), leading (in the irrotational case) to
\begin{equation}\label{bernoulli}
2 (p + gy) + u^2 + v^2 = B,
\end{equation}
where $B$ is the Bernoulli constant.
From this principle and using the mean operator \rf{defmean}, we establish some important relations, notably that $\langle\sur{u}^2+\sur{v}^2\rangle = B$ and $\langle\bot{p}\rangle = g\depth$.
Finally, we introduce the complex velocity $w(z) \eqdef u(x,y) - \ui v(x,y)$ that is holomorphic in the $z$-coordinate (with $z \eqdef x + \ui y$). 
We note that the free surface path is given by the curvilinear abscissa $\sur{z} = x + \ui\eta$ whereas the solid boundary is described by $\bot{z} = x - \ui\depth$.

The principal objective of this work is to recover the surface wave profile $\eta(x)$, the Bernoulli constant $B$ and the surface pressure $\sur{p}(x)$ from a given pressure observation $\obs{p}(x)$ at the seabed.
We assume either the wave period $L$ or the wavenumber $k$ to be a known parameter.
This inverse problem is notoriously difficult to solve due to the strong nonlinearities and its ill-posed nature (i.e. disturbances are exponentially growing to the surface).
Our first approach, developed in a series of articles \cite{C13,CC13,CH20,clamond2023steady,CLH23,LC23}, is a direct approach set in the physical space and where analytical expressions (integro-differential in general) are given in terms of the unknown surface wave profile and surface pressure.

\section{Numerical procedure: direct approach}\label{secnumapp1}

\subsection{Cauchy integral formula}

Computing steady surface waves from a boundary integral formula is an idea initiated in a seminal work by Da Silva and Peregrine \cite{DSP88}.
Hereafter, we present a similar procedure to recover an implicit expression for the surface elevation using complex analysis \cite{LC23}.
We start by giving the Cauchy integral formula, written for a holomorphic function $\Upxi(z)$, in its classical form
\begin{equation}
\label{cauchyint}
\ui\vartheta\Upxi(z) = \text{P.V.} \oint \frac{\Upxi(z')}{z'-z} \ud z' 
= \int_{-\infty}^\infty \frac{\bot{\Upxi}' \ud x'}{\bot{z}'-z} - 
\int_{-\infty}^\infty \frac{\left(1 + \ui\eta_x'\right) \sur{\Upxi}' \ud x'}
{\sur{z}'-z} ,
\end{equation}
with primes denoting the dependence on the dummy variable (e.g. $\Upxi'\eqdef\Upxi(x')$).
The argument $\vartheta$ is substituted by $\{2\pi,0,\pi\}$ when the coordinate $z$ lies respectively inside, outside and at the smooth boundary of the domain $\Omega$.

Whenever the holomorphic function is purely real at the bottom boundary (i.e. $\Im\{\bot{\Upxi}\}=0$), we may use the method of images (also called Schwarz reflection principle \cite{MorseFeshbach1953}) on \rf{cauchyint} to obtain
\begin{align}
\label{cauchypereg}
\vartheta\Upxi(z) = k \int_{\mathcal{C}} \plog{0} \{\ue^{\ui 
k (\sur{z}'-z)}\} \sur{\Upxi}' \ud\sur{z}' + k \int_{\mathcal{C}} \plog{0} 
\{\ue^{\ui k (z-\sur{\tilde{z}}'+2\ui\depth)}\} \sur{\widetilde{\Upxi}}' \ud \widetilde{\sur{z}}' + \pi \left< \sur{\Upxi} \frac{\ud \sur{z}}{\ud x} + \sur{\widetilde{\Upxi}} \frac{\ud \widetilde{\sur{z}}'}{\ud x} \right> ,
\end{align}
where $\plog{\nu}$ is the $\nu$th polylogarithm \cite{LC23} and $\widetilde{z}$ the complex conjugate of $z$.
The integration path is given (in the complex plane) by $\mathcal{C} = \{\sur{z}(x)\ :\ x\in [-\pi/k, \pi/k]\}\subset\mathbb{C}$.

Equation \eqref{cauchypereg} possesses a strong (polar) singularity causing technical difficulties when computing the integral numerically.
Thus, we use the integrated form of the polylogarithm to replace the polar term with a logarithmic singularity \cite{LC23}.
We obtain
\begin{align}
\label{dcauchypereg}
\vartheta\Upxi(z) = \ui\!\int_{\mathcal{C}} \partial_z \plog{1} \{\ue^{\ui 
k (\sur{z}'-z)}\} \sur{\Upxi}' \ud\sur{z}' + \ui\!\int_{\mathcal{C}} \partial_z \plog{1} 
\{\ue^{\ui k (z-\sur{\tilde{z}}'+2\ui\depth)}\} \sur{\widetilde{\Upxi}}' \ud \widetilde{\sur{z}}'
+ \pi \left< \sur{\Upxi} \frac{\ud \sur{z}}{\ud x} + \sur{\widetilde{\Upxi}} \frac{\ud \widetilde{\sur{z}}'}{\ud x} \right> , 
\end{align}

Computing the latter expression at the free surface reduces to
\begin{align}
\label{dcauchyperegim}
\sur{\Upxi} \ud \sur{z} = \frac{\ui}{2\pi} \ud\!\left[ \int_{\mathcal{C}} \plog{1} \{\ue^{\ui 
k (\sur{z}'-z)}\} \sur{\Upxi}' \ud\sur{z}' + \int_{\mathcal{C}} \plog{1} 
\{\ue^{\ui k (z-\sur{\tilde{z}}'+2\ui\depth)}\} \sur{\widetilde{\Upxi}}' \ud \widetilde{\sur{z}}' \right] + \frac{\ud \sur{z}}{2} \left< \sur{\Upxi} \frac{\ud \sur{z}}{\ud x} + \sur{\widetilde{\Upxi}} \frac{\ud \widetilde{\sur{z}}'}{\ud x} \right> .
\end{align}


\subsection{Expressions for computing steady surface waves and bottom pressure}

From Bernoulli's principle \eqref{bernoulli}, the complex velocity at the surface $\sur{w}$ is given explicitly by 
\begin{equation}
\label{ws}
\sur{w} = \sigma (\ud \widetilde{\sur{z}}/\ud x) \sqrt{(B-2\sur{p}-2g\eta)/\vert\ud\sur{z}/\ud x\vert^2} ,
\end{equation}
where $\sigma=\mp 1$ denotes travelling waves propagating along or against the current, respectively. 

Let us consider the holomorphic function $\Upxi=w+c$ ($c$ being an arbitrary definition of the phase speed).
The left-hand side of \rf{dcauchyperegim} follows directly from \eqref{ws} as
\begin{equation}
\label{holos}
\sur{\Upxi} \ud \sur{z} = \left(\omega h + c\right) \ud \sur{z} + \sigma \sqrt{(B-2\sur{p}-2g\eta)/\vert\ud\sur{z}/\ud x\vert^2} \ud x ,
\end{equation}
where $h\eqdef\eta+\depth$ is the wave height.
We note that the radicand is purely real since $B\geqslant 2 \vert\vert\sur{p}+g\eta\vert\vert_{\infty}$ for all waves (according to Bernoulli's relation).

After integration of expression \rf{dcauchyperegim} -- retaining the imaginary part only -- we obtain an equation for the computation of the free surface 
\begin{equation}
\label{eqeta}
\frac{\omega\/\eta^2}{2} - \eta \left< \sur{u} \left\vert\\ud\sur{z}/\ud x\right\vert^2 \right>  
 - \frac{\sigma}{2\pi} \int_{\mathcal{P}} \Re\{\mathcal{L}_{1}\} \sqrt{(B-2\sur{p}'-2g\eta')/\vert\ud\sur{z}'/\ud x'\vert^2} \ud\/x' - K = 0 ,
\end{equation}
where $K$ is a constant of integration, recovered by enforcing the mean-level condition \cite{LC23}.
For the sake of brevity, we also introduced the following notation
%
\[
\mathcal{L}_{\nu} \eqdef \plog{\nu}\!\big[\ue^{\ui k(\sur{z}'-
\sur{z})}\big] - \plog{\nu}\!\big[\ue^{\ui k(\sur{z}-
\sur{\tilde{z}}'+2\ui\depth)}\big].
\]
%
As one can notice, expression \rf{eqeta} is not dependent on the choice of phase speed $c$, as expected from the Galilean invariance of the problem.

Equation \rf{cauchypereg} and the Schwarz reflection principle also let us find an expression for the bottom velocity.
Consider $\Upxi = w$ as the holomorphic function in \rf{cauchypereg} and evaluate the expression at the bottom (with now $\vartheta=2\upi$).
Since $\bot{u}$ is a real function of the horizontal coordinate, we use the method of images to recover its explicit form, which leads to the formula
\begin{equation}
\label{ub}
\bot{u}(x) = \frac{\ui k}{4\pi} \left[ \int_{\mathcal{C}} \cot\!\left( 
k\frac{\sur{z}'-\bot{z}}{2} \right) \sur{w}' \ud \sur{z}
- \int_{\mathcal{C}} \cot\!\left( k\frac{\sur{\tilde{z}}'-\bot{z}-2\ui\depth}{2} \right) \sur{\widetilde{w}}' \ud \widetilde{\sur{z}}' \right] . 
\end{equation}

This bottom velocity $\bot{u}$ is used in Stokes' first definition of the phase speed
\begin{equation}
\label{c1}
c_1 = - \left< \bot{u} \right> .
\end{equation}

\subsection{Holomorphic functions and recovery formula}

As done numerous time in previous articles \cite{C13,CC13,CH20,clamond2023steady,CLH23,LC23}, we introduce a holomorphic \textit{complex} pressure function $\mathfrak{P}(z)$ by analytic continuation in the domain $\Omega$. 
Using Bernoulli's principle written for the complex velocity, it yields an expression for the complex pressure as
\begin{equation}
\label{P}
\mathfrak{P}(z) \eqdef g\depth + \frac{B - w^2}{2} .
\end{equation}

Similarly, we introduce the antiderivative of $\mathfrak{P}$ defined by
\begin{align} 
\label{Q}
\mathfrak{Q}(z) \eqdef \int_{z_0}^z \left[ \mathfrak{P}(z') - g\depth \right]
\ud z' = \frac{1}{2} \int_{z_0}^z \left[ B - w(z')^2 \right] \ud z', 
\end{align}
where $z_0$ is an arbitrary constant. 

In practice, these holomorphic functions are easily recovered from the given pressure measurement by fitting the data over a suitable eigenbasis (one leading to the most accuracy with the lowest order of quadrature).
Assuming a periodic motion, we fit these data with a Fourier polynomial basis before performing an analytic continuation of the pressure field.
Evaluated at the surface, it yields
\begin{equation}
\mathfrak{P}(z) = \bot{p}(z+\ui\depth)
\approx g\depth + \sum_{|n|>0}^N \mathfrak{p}_n \ue^{\ui n k (z + \ui d)} .
\end{equation}
From this definition, we simply integrate once and recover the second holomorphic function as
\begin{equation}
\sur{\mathfrak{Q}}(x) = \int_{0}^{x} \left[ \sur{\mathfrak{P}}(x') - g\depth \right] \ud x' \approx \sum_{|n|>0}^N \frac{\ui \mathfrak{p}_n}
{n k} \frac{\ue^{-nka} - \ue^{\ui nk(x+\ui\eta)}}{\ue^{nkd}} ,
\end{equation}
where $a$ is the amplitude of the wave at the crest (arbitrarily located at $x=0$).

In the context of the direct approach, we exploit a relation derived in a recent work as our general recovery formula \cite{clamond2023steady}. 
This expression (not involving any differential terms of $\eta$) is given by
\begin{equation}
\label{recoveqin}
\Re\!\left\{ \sur{\mathfrak{Q}} \right\} = \frac{1}{2} \int_{0}^{x}
\left[ B - \left|B - 2 \left( \sur{\mathfrak{P}}' - g\depth \right) \right| \right] \ud x' .
\end{equation}

As one can notice, surface pressure has been eliminated from the recovery formula \rf{recoveqin}.
This is because the Cauchy--Riemann system gives us two equations (the real and complex parts) that can be combined to suppress either the surface pressure of the surface elevation \cite{clamond2023steady}.
However, it is still necessary to fix Bernoulli's constant by adding a direct measure (phase speed, wave height, etc.) or by assuming the physics at the surface (e.g. capillary or flexural effects).
For this work, we assume Stokes's first definition of the phase speed to be a given parameter.
Thus, solving the inverse problem reduces to finding the zeros of the set of equations \rf{defmean}, \rf{c1} and \rf{recoveqin}.

\subsection{The Levenberg--Marquardt procedure}

In practice, solving the inverse problem with a direct approach is fairly straightforward.
It only requires to introduce a quadrature for numerical integration (we choose the trapezoidal quadrature) and then insert the different expressions in almost any root-finding algorithm (e.g. the \textsf{fsolve} function in \textsc{Matlab}).
Numerically, $\eta = \{\eta_i\}_i$ is a vector defined at these quadrature points.
Because this problem is notoriously challenging to solve, we decide to use a Levenberg--Marquardt algorithm, which has shown its efficiency before (see \cite{C13,CC13,CH20,clamond2023steady,CLH23,LC23}).

We quickly sketch this algorithm.
Solving the inverse problem directly (for the surface profile) can be summarised as finding the root of equation $\mathcal{E}(\eta)=0$ for some function $\mathcal{E}$.
This is equivalent as minimising the functional $\mathcal{J}(\eta) \eqdef \half \|\mathcal{E}(\eta)\|_2^2 = \half\mathcal{E} \scal \mathcal{E}$.
There are two common ways for this, both being iterative methods.
Assume we are given an initial guess $\eta^0$ with the overscript standing for the iteration step.
For any iteration $\ell$, we will denote $\mathcal{E}^\ell \eqdef \mathcal{E}(\eta^\ell)$ and $\frac{\partial \mathcal{E}^\ell}{\partial \eta} \eqdef \frac{\partial \mathcal{E}}{\partial \eta}(\eta^\ell)$ the Jacobian matrix of $\mathcal{E}$ computed at $\eta^\ell$.
\begin{itemize}
    \item[-] The first method, the Gradient Descent (GD), consists in computing the negative gradient of the function $\mathcal{J}(\eta)$ at $\eta^\ell$ in order to compute $\eta^{\ell+1}$. 
    Introducing an increment $\delta\eta_{_\text{GD}}$, expression for the gradient follows immediately
        \[-\frac{\partial \mathcal{J}}{\partial \eta}(\eta^\ell) = - \left( \frac{\partial \mathcal{E}^\ell}{\partial \eta} \right)^\intercal \mathcal{E}^\ell = \delta\eta_{_{_\text{GD}}},\]
    with $\mathcal{A}^\intercal$ denoting the transpose of a matrix $\mathcal{A}$ and where $\partial \mathcal{E}^\ell/\partial \eta$ can be approximated using finite differences. 
    This gradient is then used in an update formula as $\eta^{\ell+1} = \eta^\ell + \alpha \delta\eta_{_{_\text{GD}}}$, where $\alpha$ is the length along the descent direction.
    \item[-] The second procedure, the Gauss--Newton (GN) method, originates from a Taylor expansion of $\mathcal{E}$ at the first order. For some increment $\delta\eta_{_\text{GN}}$, it yields
        \[\begin{aligned} 
        \mathcal{J}(\eta^\ell + \delta\eta_{_\text{GN}}) \approx& \frac{1}{2}\left(\mathcal{E}^\ell + \frac{\partial \mathcal{E}^\ell}{\partial \eta} \delta\eta_{_\text{GN}}\right)\scal \left(\mathcal{E}^\ell + \frac{\partial \mathcal{E}^\ell}{\partial \eta} \delta\eta_{_\text{GN}}\right)\\
        \approx& \mathcal{J}(\eta^\ell) + \delta\eta_{_\text{GN}} \scal \left( \frac{\partial \mathcal{E}^\ell}{\partial \eta} \right)^\intercal \mathcal{E}^\ell + \frac{1}{2} \left\|\frac{\partial \mathcal{E}^\ell}{\partial \eta} \delta\eta_{_\text{GN}}\right\|_2^2.
        \end{aligned}\]
    We then minimise the function $\delta\eta_{_\text{GN}}\mapsto \mathcal{J}(\eta^\ell + \delta\eta_{_\text{GN}})$ to find the optimal increment, which is equivalent to finding the root of $\frac{\partial \mathcal{J}}{\partial \delta\eta_{_\text{GN}}}(\eta^\ell + \delta\eta_{_\text{GN}}) = 0$. After some algebraic manipulations, it gives the linear system
    \begin{equation} \label{eq:GN_update}
    \left(\frac{\partial \mathcal{E}^\ell}{\partial \eta}\right)^\intercal\frac{\partial \mathcal{E}^\ell}{\partial \eta}\delta\eta_{_\text{GN}} = -\left(\frac{\partial \mathcal{E}^\ell}{\partial \eta}\right)^\intercal\mathcal{E}^\ell = \delta\eta_{_\text{GD}}.
    \end{equation}
\end{itemize}

The Levenberg--Marquardt algorithm can be seen as an interpolation between the Gradient Descent and the Gauss--Newton method.
It consists in solving, instead of \eqref{eq:GN_update}, the Levenberg--Marquardt (LM) update as:
    \begin{equation} \label{eq:LM_update}
    \left[\left(\frac{\partial \mathcal{E}^\ell}{\partial \eta}\right)^\intercal\frac{\partial \mathcal{E}^\ell}{\partial \eta} + \lambda \mathcal{I} \right]\delta\eta_{_\text{LM}} = \delta\eta_{_\text{GD}},
    \end{equation}
where $\mathcal{I}$ is the identity matrix and $\lambda$ is a damping parameter.
A large value of $\lambda$ results in an update closer to the Gradient Descent increment, while a small value of $\lambda$ makes the update closer to the Gauss--Newton increment.
The damping parameter is updated at each iteration.
It usually starts large and becomes slower as the iterations converge towards the solution.
This procedure is used to stabilize the iterations when compared to the Gauss--Newton algorithm.
The interested reader can find a more thorough presentation of the Levenberg--Marquardt algorithm in \cite{bjorck1996numerical}.

\section{Numerical procedure: optimisation approach}\label{secnumapp2}

\subsection{An optimisation formulation}

\begin{figure}[h]
    \centering
    \begin{tikzpicture}[every text node part/.style={align=center}]
        \node[draw, text width=1.25cm] (input) at (-1,1.5) {Input $\eta$, $\sur{p}$, $B$};
        \draw[black, fill=yellow!50,text=blue] (1,1) rectangle (5,2) node[pos=.5] {Black box using \eqref{pb}};
        \node[draw, text width=1.75cm] (output) at (7.1,1.5) {Output $\bot{p}(\eta,\sur{p},B)$};
        \draw[->, >=latex] (input) -- (1,1.5);
        \draw[->, >=latex] (5,1.5) -- (output);
    \end{tikzpicture}
    \caption{Sketch of the procedure used for computing the bottom pressure from surface information.}
    \label{fig:sketch_black_box}
\end{figure}
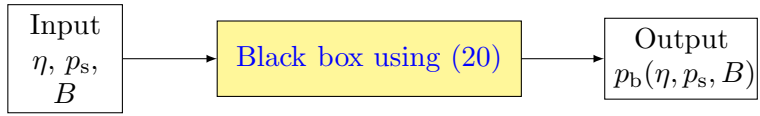

The previous approach relies on the assumption that the pressure function can be analytically continued.
Without this assumption, we can formulate the problem of recovering the surface elevation with bottom pressure measurements within an optimisation frame.
We resume the computations from \eqref{c1}.
From Bernoulli's principle, we note that given $\eta$, $\sur{p}$ and $B$, we can compute the pressure at the seabed using the formula:
\begin{equation}
\label{pb}
\bot{p}(x) = g\depth - \frac{B}{2} - \frac{k^2}{32\upi^2} \left[ \int_{\mathcal{C}} \cot\!\left( 
k\frac{\sur{z}'-\bot{z}}{2} \right) \sur{w}' \ud \sur{z}
- \int_{\mathcal{C}} \cot\!\left( k\frac{\sur{\tilde{z}}'-\bot{z}-2\ui\depth}{2} \right) \sur{\widetilde{w}}' \ud \widetilde{\sur{z}}' \right]^2 . 
\end{equation}
%
%
This procedure is sketched in figure \ref{fig:sketch_black_box} as an input--output process. 

\noindent
Ideally, we would like to find the input leading to $\bot{p}(\eta, \sur{p}, B) = \obs{p}$.
However, this strict equality may be irrelevant (due to noise in the observation for instance) and several solutions $\eta$, $\sur{p}$ and $B$ may exist (although not being physical).
Thus, we will rather minimise some distance between $\bot{p}(\eta, \sur{p}, B)$ and $\obs{p}$, while constraining the fields $\eta$, $\sur{p}$ and $B$ to respect the equality constraint \eqref{eqeta} and the mean-level condition for $\eta$.
This is summarised as the following minimisation problem
\begin{equation}\label{eq:opt_pb}
\begin{aligned}
    \min_{\eta,\sur{p}, B}\ & \mathcal{F}(\eta, \sur{p}, B) = \int_{\mathcal{P}} \left[ \bot{p}(\eta, \sur{p}, B) - \obs{p} \right]^2 \\
    \text{s.t. } & \mathcal{G}_i(\eta,\sur{p},B) = 0,\ i = 1,2,3.
 \end{aligned}
\end{equation}

where $\mathcal{G}_1$, $\mathcal{G}_2$ and $\mathcal{G}_3$ correspond to \eqref{defmean}, \eqref{eqeta} and \eqref{c1}, respectively.
As mentioned, the bottom pressure $\bot{p}$ is computed from expression \eqref{pb}.

\subsection{The augmented Lagrangian approach}

In this section, we explain the augmented Lagrangian algorithm, that we use in order to solve the problem \eqref{eq:opt_pb}.
This is an iterative method that starts with an initial guess $(\eta^0,\sur{p}^0, B^0)$ and then, based on the distance $\|\bot{p}(\eta, \sur{p}, B) - \obs{p}\|_{L^2(\mathcal{P})}^2$, will produce a new iteration $(\eta^1,\sur{p}^1, B^1)$ closer to the minimal solution.
The procedure is then iterated.
The main issue with \eqref{eq:opt_pb} are the equality constraints that should be handled with a specific procedure. 
A broad range of algorithms are specifically designed for that, such as the SQP method or the barrier methods; see \cite{boyd2004convex} for a detailed review on the numerical algorithms to solve constrained optimisation problems.
In our context, we opted for the augmented Lagrangian technique because of its ability to transform constrained problems into an unconstrained one (with a penalty term).
We introduce in the following this algorithm in a generic framework.

Given functions $\mathcal{F}$ and $\mathcal{G}$, suppose we would like to solve the generic optimisation problem
\begin{equation}
\label{eq:generic_opt}
    \min_x\ \mathcal{F}(x) \text{ s.t. }  \mathcal{G}_i(x) = 0,\ i\in\mathcal{J}.
\end{equation}
where $\mathcal{J}$ is some subset of $\mathbb{N}$.
The augmented Lagrangian algorithm offers a way to handle the constraint $\mathcal{G}(x)=0$.
It consists in taking an approximate Lagrange multipliers $\lambda^\ell_i$ (in the sense of optimisation) and coefficients $\rho^\ell_i$ to solve the optimisation problem
\[x^\ell\in \argmin_x \mathcal{F}(x) + \sum_{i\in\mathcal{J}}\lambda_i^\ell \mathcal{G}_i(x) + \sum_{i\in\mathcal{J}}\frac{\rho_i^\ell}{2}|\mathcal{G}_i(x)|^2.\]
This procedure turns the constraint problem \eqref{eq:generic_opt} into an unconstrained one, with the same cost function $\mathcal{F}$ but adding the scalar product $\lambda_i \mathcal{G}_i(x)$ and the quadratic penalization term $\half \rho_i^\ell |\mathcal{G}_i(x)|^2$.
This unconstrained minimisation problem, which will induce inner iterations, can then be solved numerically using methods for unconstrained minimisation problems ; we used the interior point method as implemented by the function \textsf{fmincon} in \textsc{Matlab} (see again \cite{boyd2004convex} for more details on this method).
Based on $x^\ell$ and for some tolerance $\mu^\ell$, the multiplier or the coefficients are updated. For each $i\in\mathcal{J}$, it follows that
\begin{itemize}
    \item[-] If $|\mathcal{G}_i(x)|\leq \mu^\ell$, then $\lambda_i^{\ell+1} = \lambda_i^\ell + \rho_i^\ell \mathcal{G}_i(x)$ and $\rho_i^{\ell+1} = \rho_i^\ell$ (the solution respects sufficiently the constraints and the multiplier is thus updated).
    \item[-] Else, $\lambda_i^{\ell+1} = \lambda_i^\ell$ and $\rho_i^{\ell+1} = \tau \rho_i^\ell$ for some $\tau>1$ (the constraints are too much violated, so we increase the weight of the quadratic penalization).
\end{itemize}
%
An analysis of this algorithm and a proof of convergence can be found in \cite{bertsekas1982constrained}.




\begin{figure}[t!]
    \centering
    \subfloat[]{
    \includegraphics*[width=.45\textwidth]{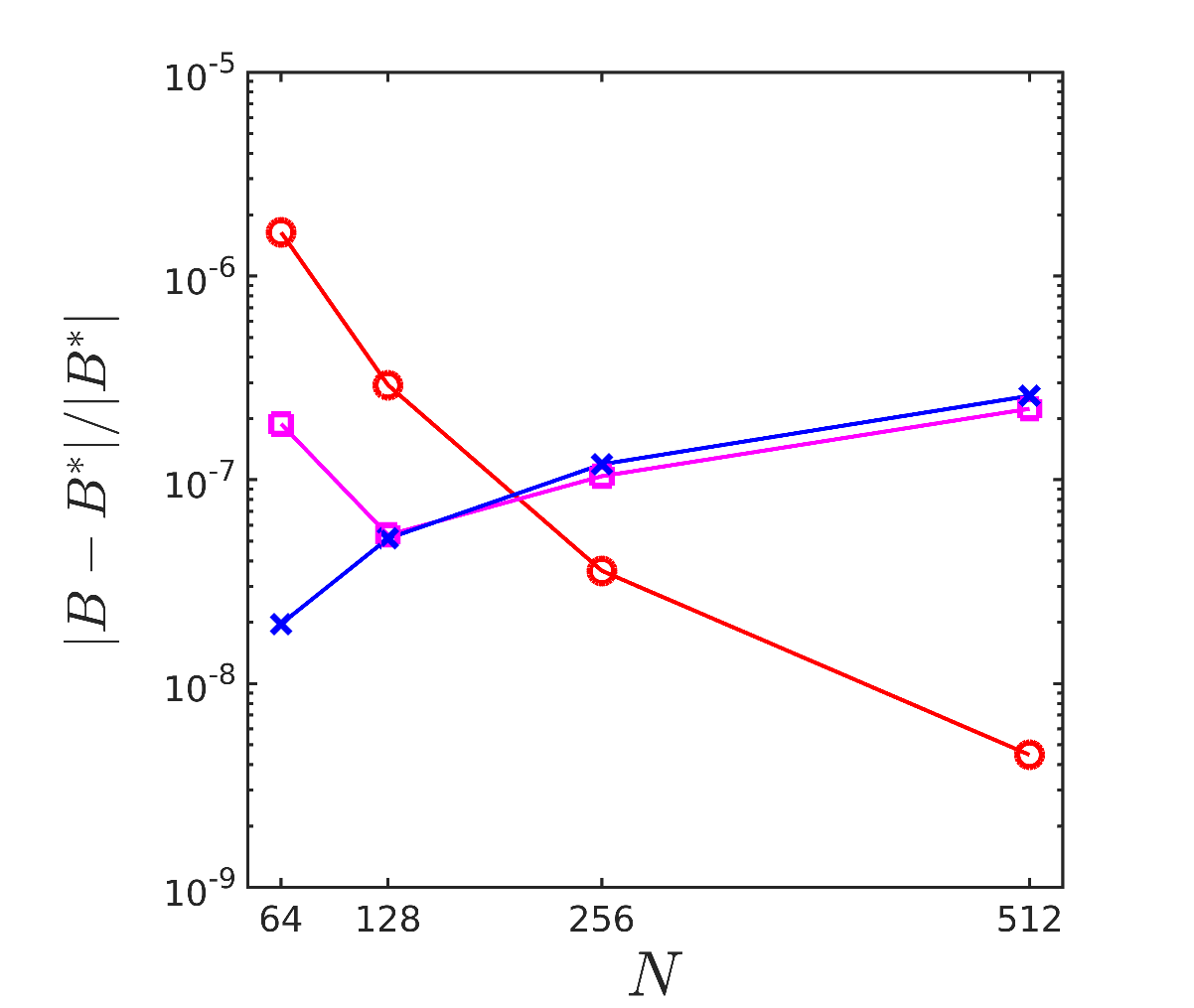} 
    \label{Fig1a}}
    \subfloat[]{
    \includegraphics*[width=.45\textwidth]{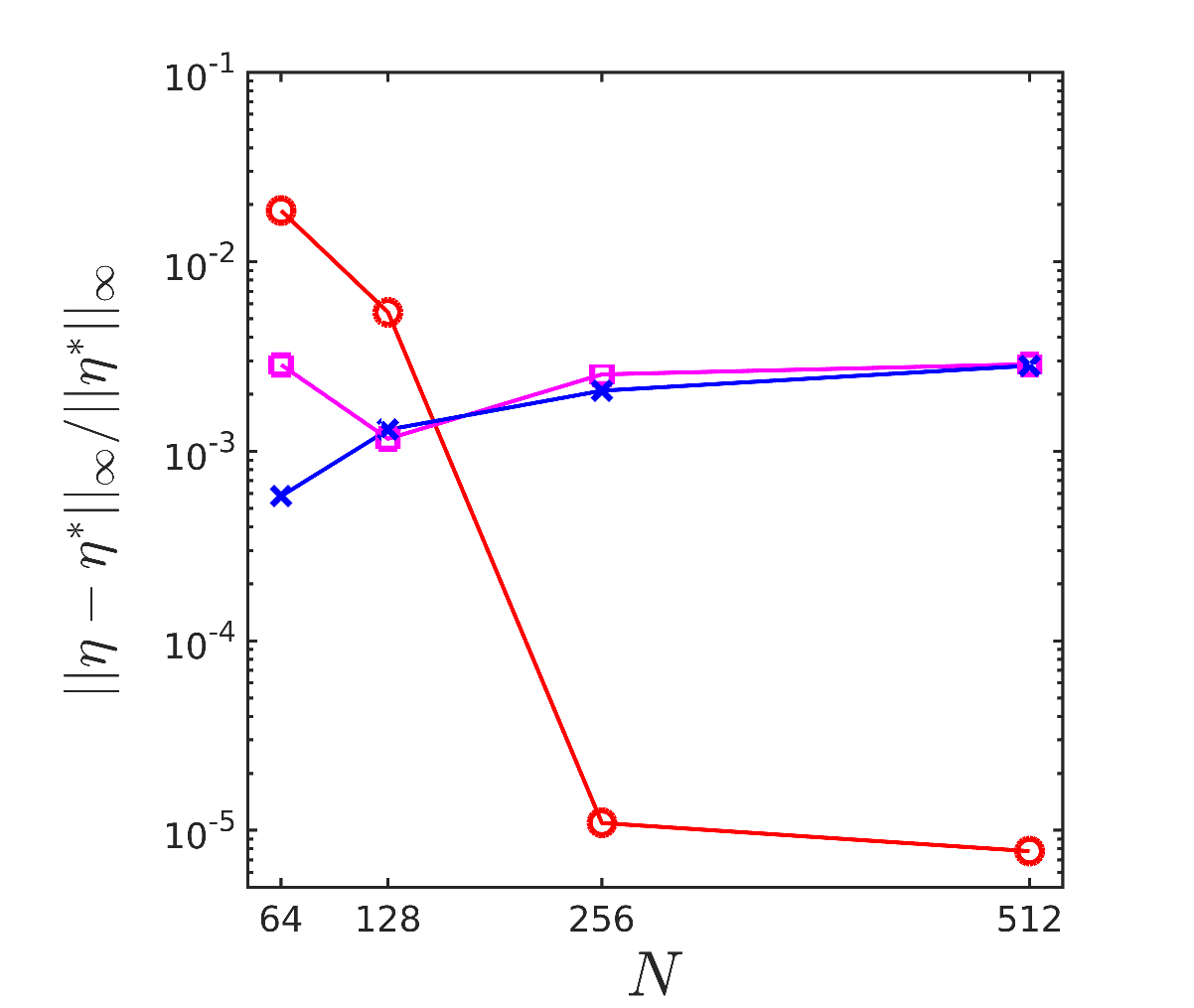} 
    \label{Fig1b}}
    \\
    \subfloat[]{
    \includegraphics*[width=.45\textwidth]{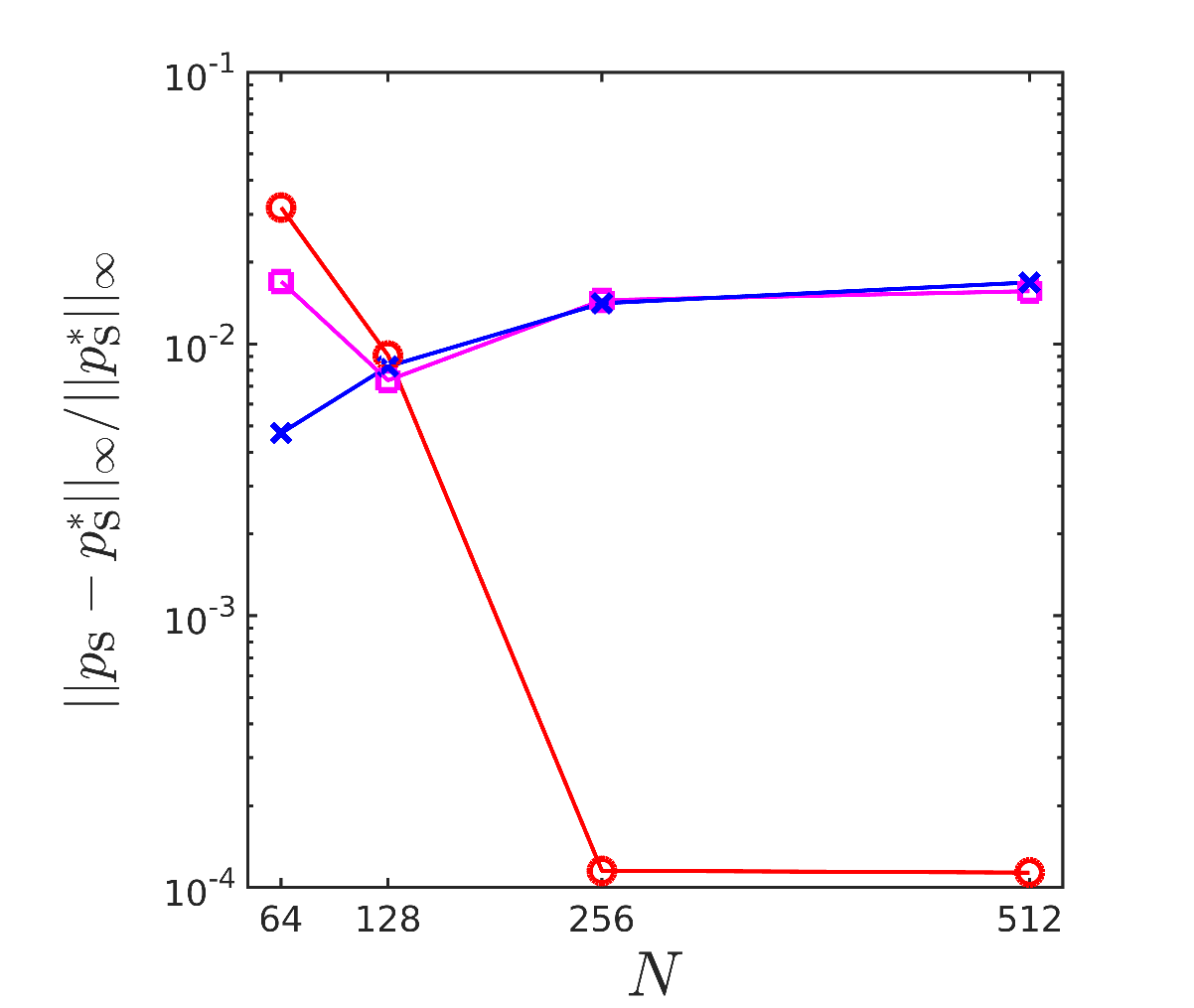} 
    \label{Fig1c}}
    \subfloat[]{
    \includegraphics*[width=.45\textwidth]{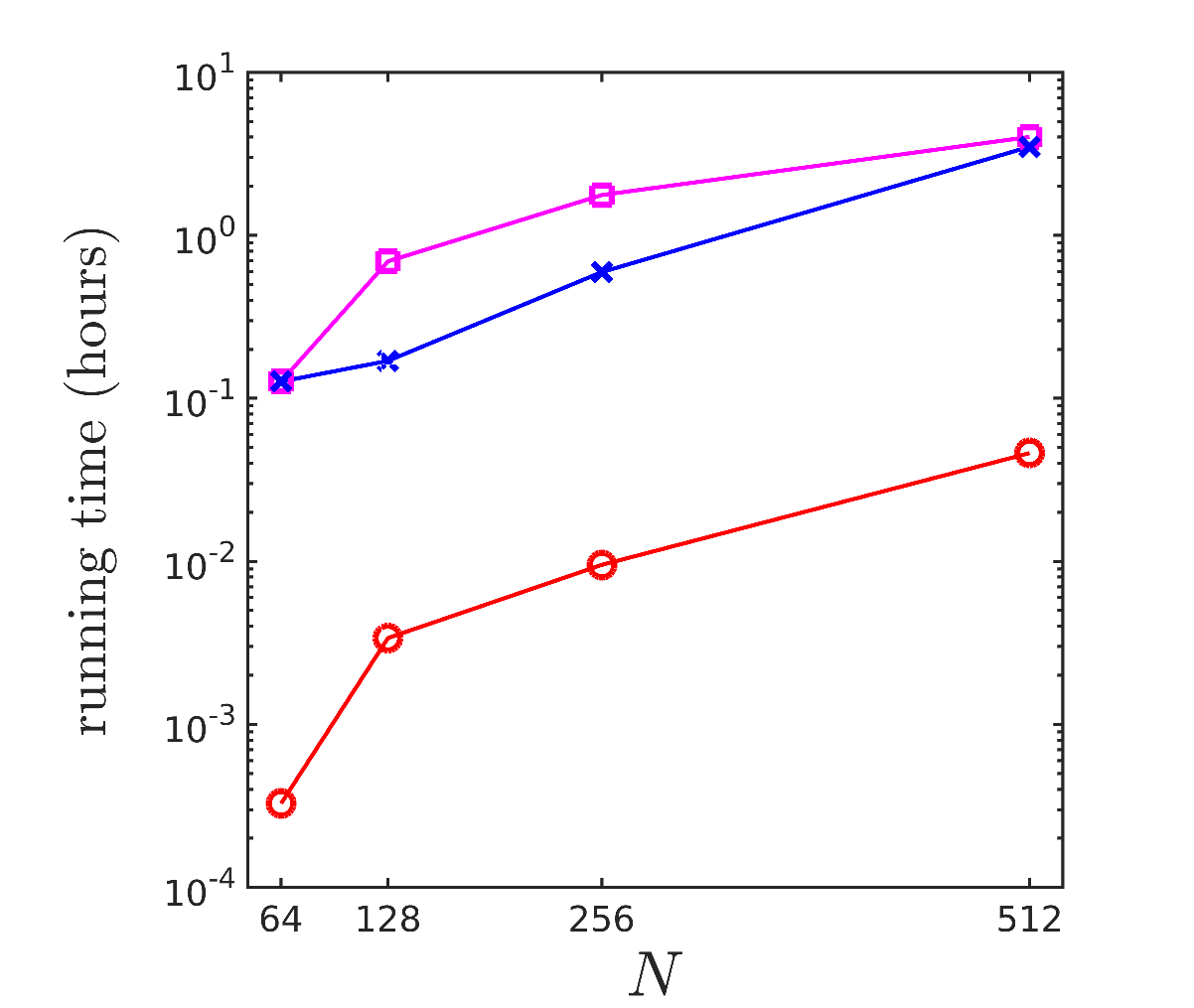} 
    \label{Fig1d}}
    \caption{Relative errors versus the number of Fourier modes $N$ for (a): $B$, (b): $\eta$ and (c): $\sur{p}$. We give the running time (in hours) in panel (d). The circled red lines correspond to the direct approach. The crossed blue (squared magenta) lines represent the solutions from the optimisation method with a low (high) limit on the number of function evaluations.}
    \label{Fig1}
\end{figure}

\section{Numerical experiments}\label{secnumexp}

We now test numerically the two methods (direct and optimisation) on a common example of surface profile recovery.
This example was already treated in \cite{clamond2023steady} and appeared as one of the most difficult case considered.
We mention that every computations were made on a processor AMD EPYC 7542 @1.5Ghz with 512GB of RAM.

\subsection{Recovery of capillary waves}

When considering a fluid with surface tension $\gamma$, the analytic form of the pressure at the surface is given by the formula
\begin{equation} \label{eq:ps}
\sur{p} \eqdef -\gamma\kappa = -\frac{\gamma\eta_{xx}}{\left(1+\eta_x^2\right)^{3/2}} ,
\end{equation}
where $\kappa$ is the curvature of the interface.
It was shown recently \cite{clamond2023steady} that in this context, recovery of the surface elevation is a challenging task due to the presence of high-order derivatives in the nonlinear terms.
Nevertheless, we demonstrate the efficiency of both approaches by considering a gravity-capillary wave with $\gamma=1/3$, $\sigma=-1$ and $L/d=6\pi$.
The solutions of reference $B^{*}$, $\eta^{*}$ and $\sur{p}^{*}$ are computed from an algorithm adapted from \cite{LC23} with the total wave height being fixed at $H/\depth=0.1$.
For the initial guess, we use the solutions given by linear theory for both numerical approaches.
In addition, for the Levenberg--Marquardt algorithm, we used an initial damping parameter set to $\lambda=0.05$.
The initial parameters for the optimisation algorithm are set to $\lambda^0_i=0$, $\rho^0_i=10$ and $\tau^0=2$. 

First, we let vary the number of Fourier modes $N$ and observe the eventual convergence of the solutions.
In figure \ref{Fig1}, we display with circled red lines the relative errors on $B$, $\eta$ and $\sur{p}$, as well as the running time, for the direct approach.
We superpose the results from the optimisation approach with a low (high) limit on the number of function evaluations in the inner iterations using crossed blue (squared magenta) lines.
All these data are summarised in the rows of table \ref{tab:errors} without noise.
As one can notice in figure \ref{Fig1}, the accuracy of the direct approach continues to increase with $N$ whereas the errors from the optimisation method slightly grow.
However, restricting the number of function evaluations on the latter allows for faster (and more accurate) results when the number of modes is small enough.
Finally, we notice that the computation time is in favor of the direct approach by far ($\sim100$ times faster).

\begin{figure}[t!]
    \centering
    \subfloat[]{
    \includegraphics*[width=.95\textwidth]{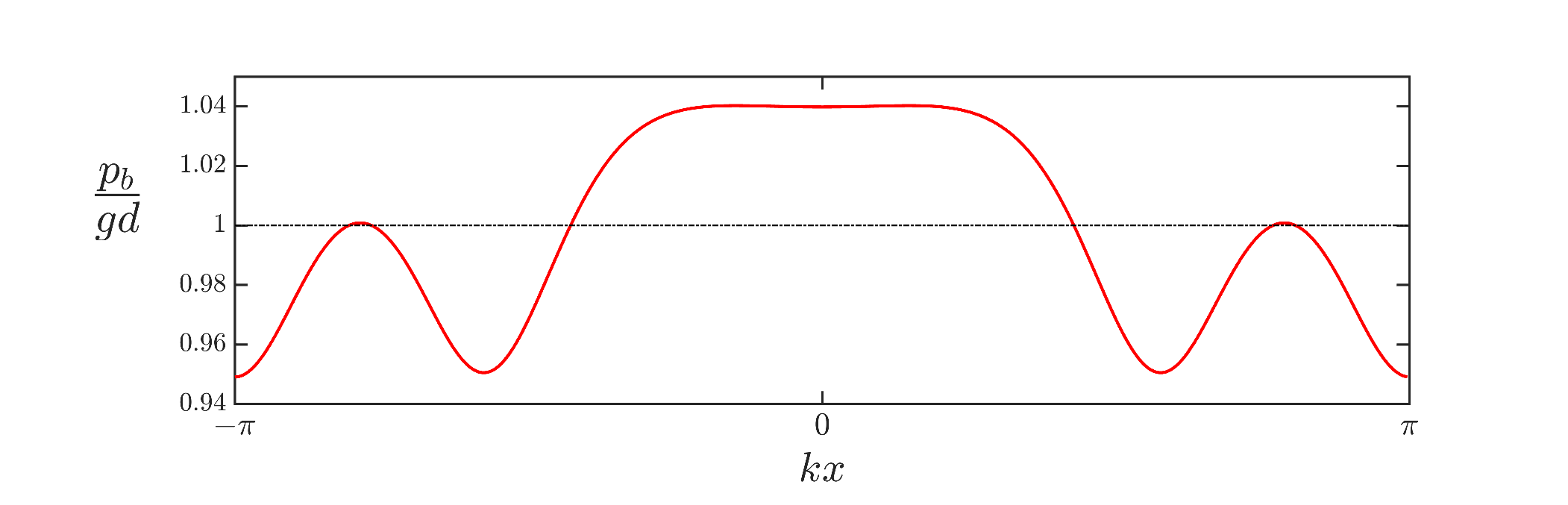} 
    \label{Fig2a}}
    \vspace{-.3em}\\
    \subfloat[]{
    \includegraphics*[width=.95\textwidth]{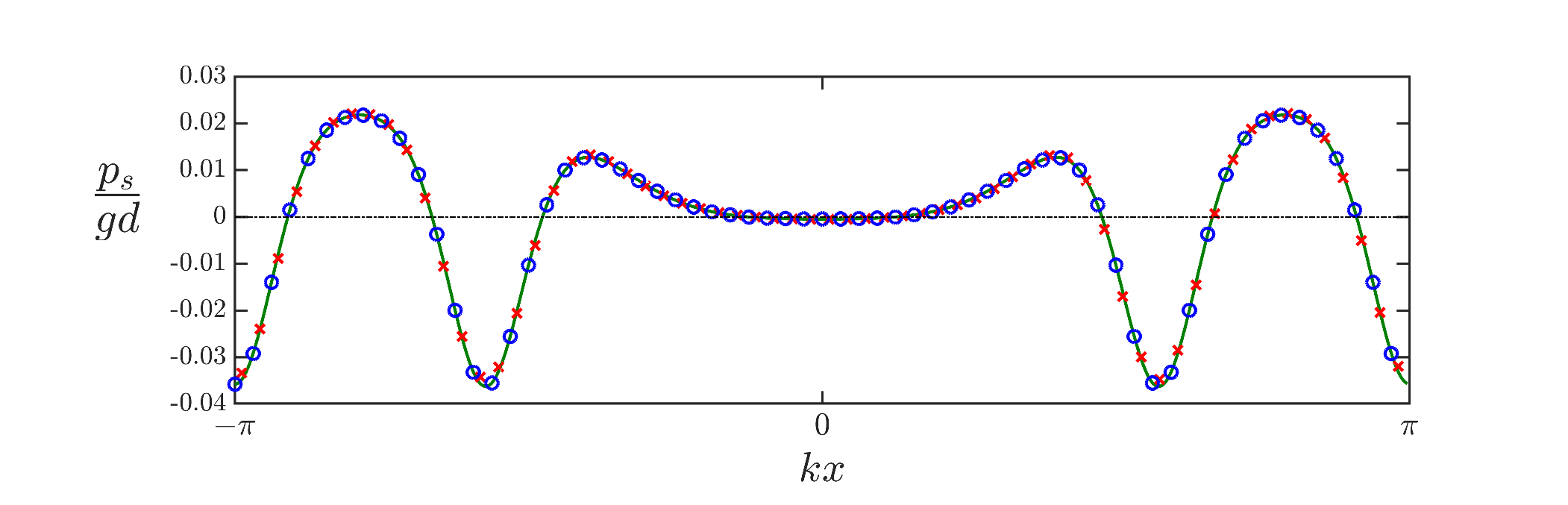} 
    \label{Fig2b}}
    \vspace{-.3em}\\
    \subfloat[]{
    \includegraphics*[width=.95\textwidth]{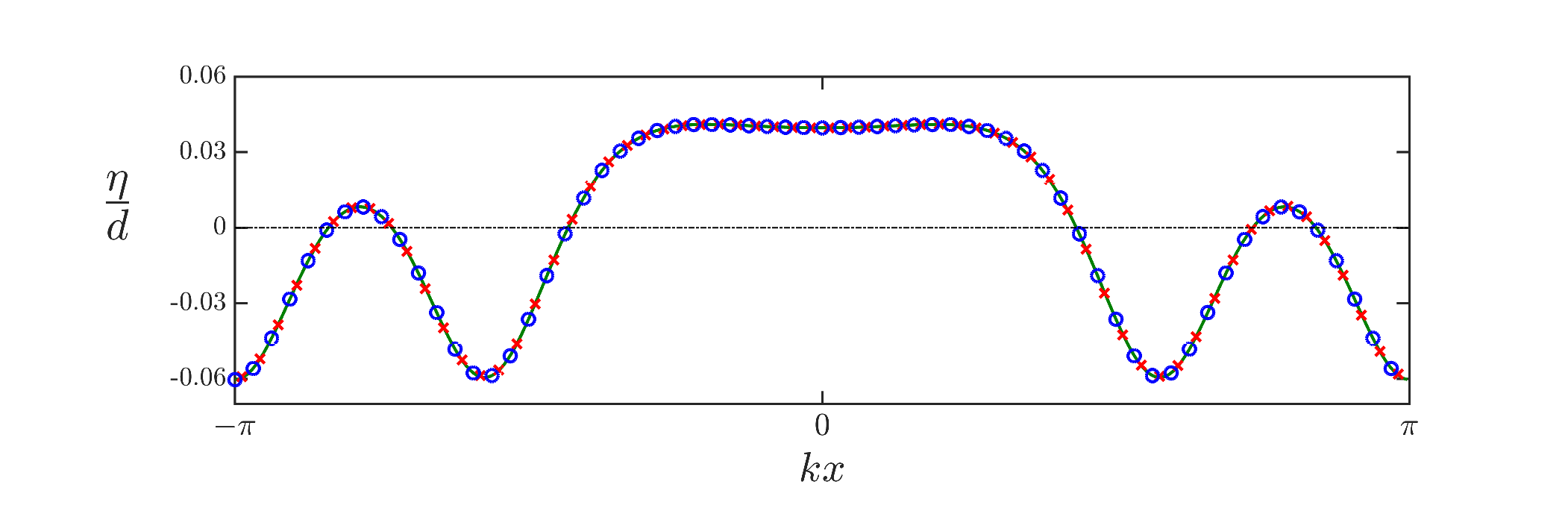} 
    \label{Fig2c}}
    \caption{(a): Bottom pressure of reference, computed from expression \rf{pb}. The dashed-dotted black line represents the hydrostatic value $g\depth$. (b,c): Recovered surface pressure and surface elevation from direct approach (red crosses) and optimisation approach (blue circles). The reference values are displayed in dark green lines. All computations were made with $N=512$.}
    \label{Fig2}
\end{figure}

Eventually, we compare the recovered solutions and the reference values in figure \ref{Fig2}.
Starting from the bottom pressure in panel \ref{Fig2a}, we obtain the surface elevation and the surface pressure by direct approach (red crosses) or by optimisation method (blue circles).
Both results are in excellent agreement with the reference value given by the dark green curves.

\subsection{Addition of noise in the measurement}

We now test the stability of the method to random perturbations of the observation signal.
For this, we add a white noise to the pressure of reference $\obs{p}$ and try to recover solutions at the surface and Bernoulli's constant from this noisy data.
This is of significant importance from an engineering point of view, since most measurements will necessarily contain noise.

\begin{figure}[t!]
    \centering
    \subfloat[]{
    \includegraphics*[width=.95\textwidth]{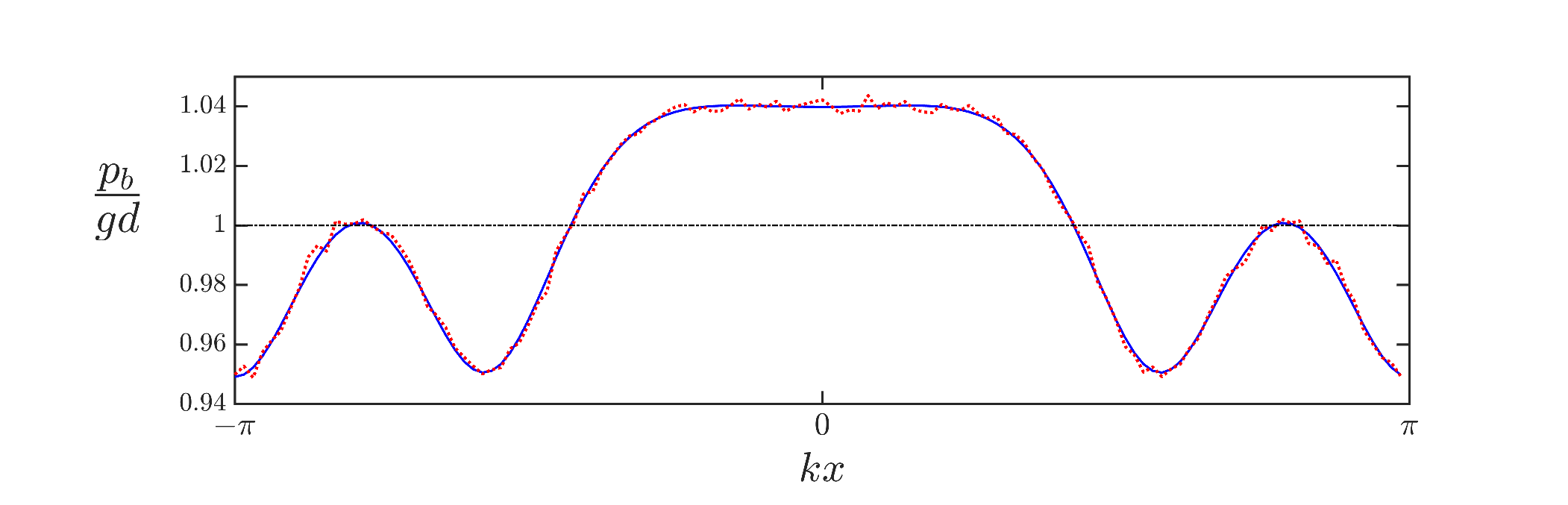} 
    \label{Fig3a}}
    \vspace{-.3em}\\
    \subfloat[]{
    \includegraphics*[width=.95\textwidth]{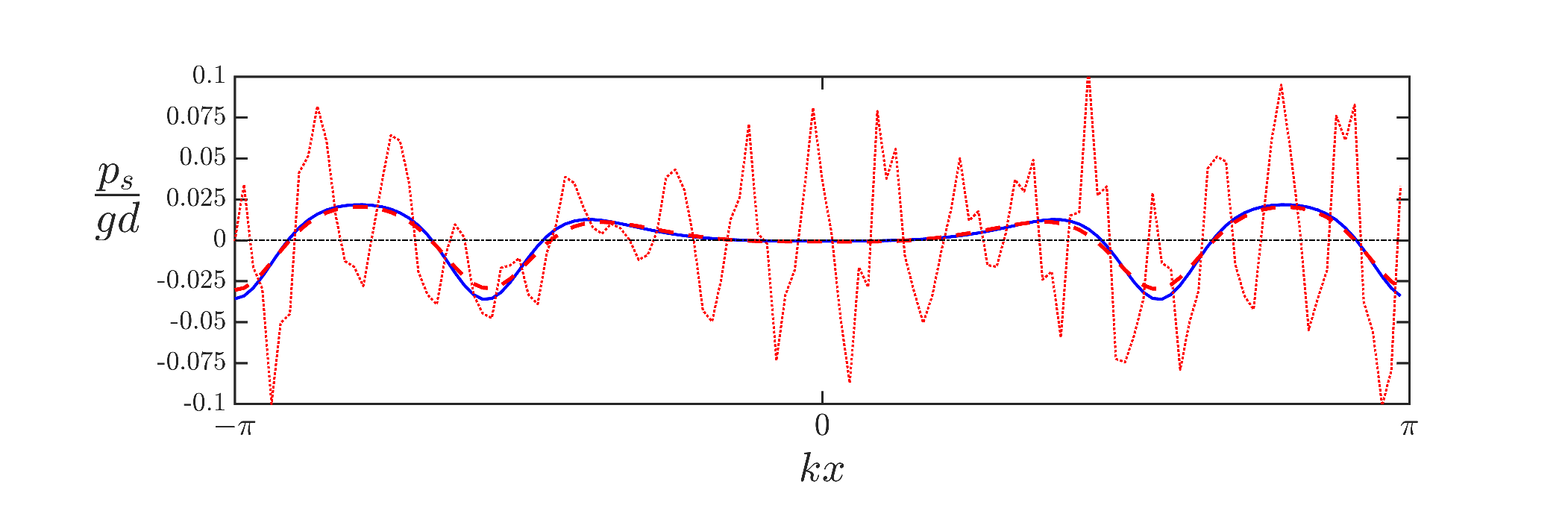} 
    \label{Fig3b}}
    \vspace{-.3em}\\
    \subfloat[]{
    \includegraphics*[width=.95\textwidth]{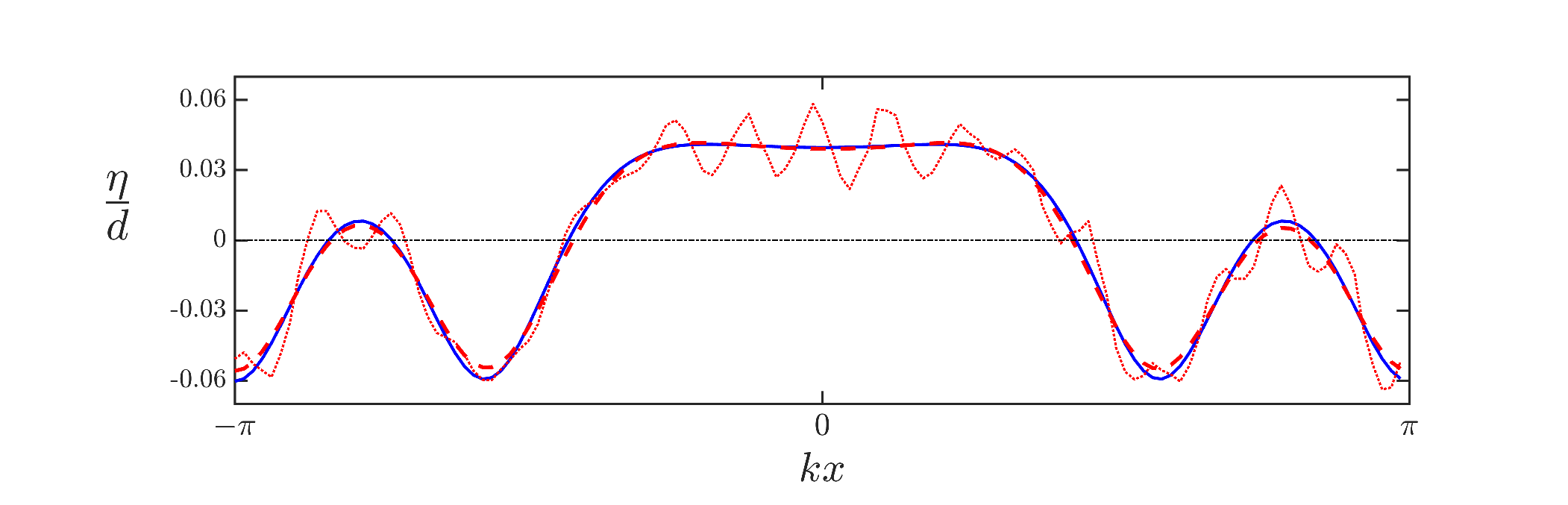} 
    \label{Fig3c}}
    \caption{(a): Pressure at the seabed. Ground truth value (computed from \rf{pb}) is depicted in blue line. The signal with $5\%$ added noise is represented in dotted red line. (b,c): Recovered surface pressure and surface elevation without (with) knowledge of the physics at the surface are represented in dotted (dashed) red lines. The reference values are displayed with blue lines. All computations were made with $N=128$.}
    \label{Fig3}
\end{figure}

Starting from the pressure field $\obs{p}$ without noise, we create a noisy signal $\obs{\widecheck{p}} = \obs{p} + \varepsilon$. 
Here, the white noise $\varepsilon \sim \mathcal{N}(0, \text{STD})$ is designed using the standard deviation $\text{STD} = \frac{1}{3} \epsilon (\max(\obs{p})-\min(\obs{p}))$, where $\epsilon$ will be some prescribed error. 
This let us have an error on the signal which is at most $\epsilon (\max(\obs{p})-\min(\obs{p}))$ with a 99\% confidence.
In our experiments, we use $\epsilon = 1\%, 5\%$ and $10\%$.
All results are shown in Table \ref{tab:errors}.
The errors should be understood as the relative error (with respect to the sup norm) to the expected solutions without noise.
As deduced from these results, the direct approach fails to recover the surface even with the smallest amount of noise added.
In contrast, the optimisation process produces results in better agreement with the solution sought.
This can also be emphasised from figure \ref{Fig3}, where we used $N=128$ modes and $\epsilon = 5\%$.
We notice that all errors are of the same magnitude as the noise added to the original signal, exception made of $\sur{p}$.
However, the computation time needed to solve the inverse problem also grows significantly although it is not proportional to the rate of noise added.

\input{dat_folder/input_tables}

\section{Discussion}\label{secconclu}

We now discuss the results presented above and draw some conclusions from them.
We will mainly discuss two aspects: the effect of noise addition on the two methods, and the comparison of the direct and optimisation methods from a technical perspective.
Both topics will be the occasion to hint some possible extensions and future research directions.

\subsection{Comments on the noise addition}

Implementing noise in our numerical experiments provided some evidence that analytical continuation of the bottom pressure is a prerequisite for the direct method to work.
Even in the case with only 1\% of error, results in table \ref{tab:errors} show that it renders this method dramatically ineffective.
In contrast, the optimisation procedure shows a good agreement even in the presence of noise.
For instance, the solution fields $\eta$ and $B$ have a relative error of the same magnitude as the noise added, which demonstrates the robustness of this method.
Actually, this observation seems natural.
The way we formulate the inverse problem using an optimisation frame can be seen as a filtering for the bottom pressure, since minimizing the $L_2$ distance between $\bot{p}(\eta, \sur{p}, B)$ and $\obs{\widecheck{p}}$ corresponds to fitting $\bot{p}$ to some unbiased estimator of $\obs{\widecheck{p}}$.
However, the surface pressure $\sur{p}$ still shows a non-negligible deviation from the sought solution.
As mentioned in the introduction, this can be attributed to the ill-posedness of the problem, namely that any disturbances in the measurements propagate exponentially to the surface.
This is further explained by the nature of the dynamic condition at the surface and the physical assumptions on $\sur{p}$.
Indeed, capillary (or flexural) waves involve high-order derivatives of the surface profile that are numerically problematic to approximate when the solution is not smooth enough (as it is the case with unfiltered noise).
Numerical errors can thus accumulate since there are no constraint to enforce any regularity on the surface pressure.

We see several ways to accommodate both methods to the case where noise is added.
Regarding the direct approach, we should focus on filtering the noise in the bottom measurements before solving the analytical expressions.
This method would require a precise estimation technique to guarantee the analyticity of the signal, since our experiments show that even the slightest addition of noise breaks the whole recovery process.
As for the optimisation method, the main issue in the presence of noise solely concerns the surface pressure $\sur{p}$.
This can be solved by including more information to the problem, e.g. some additional measurements or some knowledge of the physical mechanisms (the two being not exclusive).
In practice, this added information is simply incorporated in the optimisation problem by adding an extra constraint.
However, the latter could make the problem unsolvable or at least numerically harder to solve.
We illustrate this idea with a numerical experiment on the same configuration than before where we added the constraint $\|\sur{p}-\sur{p}^{\textrm{th}}(\eta)\|^2_{L^2(\mathcal{P})}=0$ in the optimisation problem \eqref{eq:opt_pb} ($\sur{p}^{\textrm{th}}$ being computed from the analytic expression \eqref{eq:ps}).
The results for $N=128$ and $\epsilon=5\%$ are depicted in figure \ref{Fig3} with a dashed red curve when adding this new constraint and in dotted red line without it.
As clear from these panels, knowing the physics at the surface removes the oscillations due to the noise and allows to recover the solutions of reference with an excellent agreement.
However, this configuration becomes extremely challenging to solve when considering this additional constraint, as clear from the computation time taken to solve this problem (which was $t\sim5.1243$ hours).
Future research should focus on the nature of information one can add to stabilise the solutions and lower the complexity.

\subsection{Comparison of the direct and optimisation approaches}

The principal objective of this work was to solve a nonlinear ill-posed inverse problem using two distinct approaches and compare these methods from a numerical perspective.
At first glance, we notice a clear discrepancy in the computation time of both procedures.
The direct approach is indeed much faster than the optimisation method and we would like to understand the reasons behind this.
Let us return to the intrinsic nature of our direct formulation.
By considering the analytic extension of the bottom pressure, it allows for the derivation of some implicit formula for the recovery of the surface profile.
Solving these expressions with a root-finding algorithm represents an efficient way of solving this inverse problem with a system of reduced size.
Therefore, the direct approach naturally exhibits one of the fastest way to compute these variables with an excellent accuracy.
On the other hand, the optimisation procedure \rf{eq:opt_pb} requires to solve a series of unconstrained problem iteratively and adjust the parameters at each iteration to respect the constraints.
This method acts on a system with a larger size and is strongly dependent on the number of constraints and optimisation variables.
Indeed, we observe in figure \ref{Fig1} (with data summarised in table \ref{tab:errors} with $0\%$ error) a tendency of the errors on the recovered variables to grow as the size of the system increases, while the direct method takes advantage of this increased number of variables.
This is most probably due to the complexity of the formulation (i.e. respecting all the constraints simultaneously) becoming too arduous.
Acceleration of the optimisation method should be an important topic of research in order to make this approach really efficient and could be made using different techniques, such as an increased parallelisation of the computation or via techniques of distributed optimisation ; see \cite{bertsekas2015parallel, boyd2011distributed} for more details on this topic. 

Moreover, both methods are interesting in the sense that they can be generalised fairly easily to more complex situations.
Adding a linear shear current (as done in \cite{CLH23}) and considering overhanging profiles (as done in \cite{LC23}) or flexural effects at the surface is straightforward to implement in both approaches, although it would increase the complexity of the expressions we used.
Extending this work to the case of unsteady motions is a challenging task that is still lacking from the literature. 
It would require in practice to solve the inverse problem iteratively in time (repeating the current technique at each time step) while avoiding numerical instabilities (e.g. aliasing) and following the same branch of solution.
In principle, both methods are applicable when considering this configuration although the running time of the optimisation problem represents a clear impediment to this formulation.
Finally, extension towards the three-dimensional case is most certainly the principal reason why we considered the optimisation approach in the first place.
This formulation is indeed well-designed to handle three-dimensional problems ; see \cite{challis2009level, gaymann2019fluid, villanueva2017cutfem} for examples of optimisation problems involving 3D equations. 
However, it would require to solve the equations of motion directly instead of considering the analytical expressions we used in this work that were obtained from a boundary integral method.
Since these expressions are expressed in the complex plane, their generalisation in 3D is not trivial and represents a current topic of research.
For this reason, the direct approach is less inclined to be extended to this configuration whereas the optimisation formulation could present a solution to this long-standing problem.\\


\noindent{\bf Funding.} Joris Labarbe has been supported by the French government, 
through the $\mbox{UCA}^{\mbox{\tiny JEDI}}$ {\em Investments in the Future\/} 
project managed by the National Research Agency (ANR) with the reference number 
ANR-15-IDEX-01. \\

\noindent{\bf Declaration of interests.} The authors report no conflict of interest.

\bibliographystyle{amsplain}
\bibliography{Biblio}

\end{document}

%% file: dat_folder/input_tables.tex
\begin{table}[!t]
    \begin{subtable}[h]{0.45\textwidth}
        \centering
        \begin{tabular}{|c|c|m{5em}|m{5em}|}%
            \hline
            \bfseries \% noise & \bfseries N & \bfseries Direct & \bfseries Optim.\\
            \hline
            \begin{filecontents*}{dat_folder/err_Bs.dat}
perErr,N,Bo,Bd
0,64,1.9636e-08,1.6407e-06
0,128,5.1596e-08,2.9192e-07
0,256,1.1931e-07,3.5804e-08
0,512,2.5856e-07,4.4756e-09
1,64,2.4425e-05,5.4159e+00
1,128,1.0315e-04,1.6627e+00
1,256,3.8279e-04,6.9348e-01
5,64,8.7488e-04,2.1112e+01
5,128,2.2192e-04,1.1758e+01
5,256,1.4831e-04,6.6573e+00
10,64,9.1691e-03,9.7216e+00
10,128,2.0514e-04,1.0309e+01
10,256,2.2797e-03,1.0938e+01
\end{filecontents*}
\csvreader[
            head to column names, 
            filter expr = { test{\ifnumequal{\perErr}{0}}
                            and test{\ifnumequal{\N}{64}} },
            ]
            {dat_folder/err_Bs.dat}{}
            {\multirow{4}{*}{\perErr} & \N & \Bd & \Bo }
            \csvreader[
            head to column names, 
            filter expr = { test{\ifnumequal{\perErr}{0}}
                            and test{\ifnumgreater{\N}{65}} },
            ]
            {dat_folder/err_Bs.dat}{}
            {\\ \cline{2-4}
            & \N & \Bd & \Bo}
            \\
            
            \hline
            \csvreader[
            head to column names, 
            filter expr = { test{\ifnumequal{\perErr}{1}}
                            and test{\ifnumequal{\N}{64}} },
            ]
            {dat_folder/err_Bs.dat}{}
            {\multirow{3}{*}{\perErr} & \N & \Bd & \Bo }
            \csvreader[
            head to column names, 
            filter expr = { test{\ifnumequal{\perErr}{1}}
                            and test{\ifnumgreater{\N}{65}} },
            ]
            {dat_folder/err_Bs.dat}{}
            {\\ \cline{2-4}
            & \N & \Bd & \Bo}
            \\
            
            \hline
            \csvreader[
            head to column names, 
            filter expr = { test{\ifnumequal{\perErr}{5}}
                            and test{\ifnumequal{\N}{64}} },
            ]
            {dat_folder/err_Bs.dat}{}
            {\multirow{3}{*}{\perErr} & \N & \Bd & \Bo }
            \csvreader[
            head to column names, 
            filter expr = { test{\ifnumequal{\perErr}{5}}
                            and test{\ifnumgreater{\N}{65}} },
            ]
            {dat_folder/err_Bs.dat}{}
            {\\ \cline{2-4}
            & \N & \Bd & \Bo}
            \\
            
            \hline
            \csvreader[
            head to column names, 
            filter expr = { test{\ifnumequal{\perErr}{10}}
                            and test{\ifnumequal{\N}{64}} },
            ]
            {dat_folder/err_Bs.dat}{}
            {\multirow{3}{*}{\perErr} & \N & \Bd & \Bo }
            \csvreader[
            head to column names, 
            filter expr = { test{\ifnumequal{\perErr}{10}}
                            and test{\ifnumgreater{\N}{65}} },
            ]
            {dat_folder/err_Bs.dat}{}
            {\\ \cline{2-4}
            & \N & \Bd & \Bo}
            \\
            \hline
        \end{tabular}
    \caption{$|B-B^*|/|B^*|$.}
    \label{tab:err_B}
    \end{subtable}
    \hspace{1em}
    \begin{subtable}[h]{0.45\textwidth}
        \centering
        \begin{tabular}{|c|c|m{5em}|m{5em}|}%
            \hline
            \bfseries \% noise & \bfseries N & \bfseries Direct & \bfseries Optim.\\
            \hline
            \begin{filecontents*}{dat_folder/err_ps.dat}
perErr,N,pso,psd
0,64,4.6998e-03,3.1851e-02
0,128,8.2738e-03,9.0898e-03
0,256,1.4128e-02,1.1527e-04
0,512,1.6856e-02,1.1342e-04
1,64,9.5117e-01,9.0614e+01
1,128,1.5885e+00,3.5142e+01
1,256,3.3760e+00,2.2646e+01
5,64,4.8191e+00,1.5840e+03
5,128,2.7363e+00,4.3467e+02
5,256,2.2819e+00,2.4527e+03
10,64,6.7018e+00,7.1252e+03
10,128,4.1020e+00,1.9906e+04
10,256,9.1385e+00,3.1396e+04
\end{filecontents*}
\csvreader[
            head to column names, 
            filter expr = { test{\ifnumequal{\perErr}{0}}
                            and test{\ifnumequal{\N}{64}} },
            ]
            {dat_folder/err_ps.dat}{}
            {\multirow{4}{*}{\perErr} & \N & \psd & \pso }
            \csvreader[
            head to column names, 
            filter expr = { test{\ifnumequal{\perErr}{0}}
                            and test{\ifnumgreater{\N}{65}} },
            ]
            {dat_folder/err_ps.dat}{}
            {\\ \cline{2-4}
            & \N & \psd & \pso}
            \\
            
            \hline
            \csvreader[
            head to column names, 
            filter expr = { test{\ifnumequal{\perErr}{1}}
                            and test{\ifnumequal{\N}{64}} },
            ]
            {dat_folder/err_ps.dat}{}
            {\multirow{3}{*}{\perErr} & \N & \psd & \pso }
            \csvreader[
            head to column names, 
            filter expr = { test{\ifnumequal{\perErr}{1}}
                            and test{\ifnumgreater{\N}{65}} },
            ]
            {dat_folder/err_ps.dat}{}
            {\\ \cline{2-4}
            & \N & \psd & \pso}
            \\
            
            \hline
            \csvreader[
            head to column names, 
            filter expr = { test{\ifnumequal{\perErr}{5}}
                            and test{\ifnumequal{\N}{64}} },
            ]
            {dat_folder/err_ps.dat}{}
            {\multirow{3}{*}{\perErr} & \N & \psd & \pso }
            \csvreader[
            head to column names, 
            filter expr = { test{\ifnumequal{\perErr}{5}}
                            and test{\ifnumgreater{\N}{65}} },
            ]
            {dat_folder/err_ps.dat}{}
            {\\ \cline{2-4}
            & \N & \psd & \pso}
            \\
            
            \hline
            \csvreader[
            head to column names, 
            filter expr = { test{\ifnumequal{\perErr}{10}}
                            and test{\ifnumequal{\N}{64}} },
            ]
            {dat_folder/err_ps.dat}{}
            {\multirow{3}{*}{\perErr} & \N & \psd & \pso }
            \csvreader[
            head to column names, 
            filter expr = { test{\ifnumequal{\perErr}{10}}
                            and test{\ifnumgreater{\N}{65}} },
            ]
            {dat_folder/err_ps.dat}{}
            {\\ \cline{2-4}
            & \N & \psd & \pso}
            \\
            \hline
        \end{tabular}
    \caption{$\|\sur{p}-\sur{p}^*\|_\infty/\|\sur{p}^*\|_\infty$.}
    \label{tab:err_ps}
    \end{subtable}
    \\
    \begin{subtable}[h]{0.45\textwidth}
        \centering
        \begin{tabular}{|c|c|m{5em}|m{5em}|}%
            \hline
            \bfseries \% noise & \bfseries N & \bfseries Direct & \bfseries Optim.\\
            \hline
            \begin{filecontents*}{dat_folder/err_eta.dat}
perErr,N,etao,etad
0,64,5.8043e-04,1.8648e-02
0,128,1.3014e-03,5.4193e-03
0,256,2.0904e-03,1.0942e-05
0,512,2.8299e-03,7.7705e-06
1,64,1.5490e-01,2.7749e+00
1,128,1.2423e-01,2.6303e+00
1,256,8.4355e-02,2.5163e+00
5,64,1.0194e+00,3.9779e+00
5,128,3.1647e-01,3.7786e+00
5,256,1.3457e-01,6.4795e+00
10,64,3.0907e+00,5.3201e+00
10,128,5.4415e-01,8.6517e+00
10,256,3.5418e-01,9.7605e+00
\end{filecontents*}
\csvreader[
            head to column names, 
            filter expr = { test{\ifnumequal{\perErr}{0}}
                            and test{\ifnumequal{\N}{64}} },
            ]
            {dat_folder/err_eta.dat}{}
            {\multirow{4}{*}{\perErr} & \N & \etad & \etao }
            \csvreader[
            head to column names, 
            filter expr = { test{\ifnumequal{\perErr}{0}}
                            and test{\ifnumgreater{\N}{65}} },
            ]
            {dat_folder/err_eta.dat}{}
            {\\ \cline{2-4}
            & \N & \etad & \etao}
            \\
            
            \hline
            \csvreader[
            head to column names, 
            filter expr = { test{\ifnumequal{\perErr}{1}}
                            and test{\ifnumequal{\N}{64}} },
            ]
            {dat_folder/err_eta.dat}{}
            {\multirow{3}{*}{\perErr} & \N & \etad & \etao }
            \csvreader[
            head to column names, 
            filter expr = { test{\ifnumequal{\perErr}{1}}
                            and test{\ifnumgreater{\N}{65}} },
            ]
            {dat_folder/err_eta.dat}{}
            {\\ \cline{2-4}
            & \N & \etad & \etao}
            \\
            
            \hline
            \csvreader[
            head to column names, 
            filter expr = { test{\ifnumequal{\perErr}{5}}
                            and test{\ifnumequal{\N}{64}} },
            ]
            {dat_folder/err_eta.dat}{}
            {\multirow{3}{*}{\perErr} & \N & \etad & \etao }
            \csvreader[
            head to column names, 
            filter expr = { test{\ifnumequal{\perErr}{5}}
                            and test{\ifnumgreater{\N}{65}} },
            ]
            {dat_folder/err_eta.dat}{}
            {\\ \cline{2-4}
            & \N & \etad & \etao}
            \\
            
            \hline
            \csvreader[
            head to column names, 
            filter expr = { test{\ifnumequal{\perErr}{10}}
                            and test{\ifnumequal{\N}{64}} },
            ]
            {dat_folder/err_eta.dat}{}
            {\multirow{3}{*}{\perErr} & \N & \etad & \etao }
            \csvreader[
            head to column names, 
            filter expr = { test{\ifnumequal{\perErr}{10}}
                            and test{\ifnumgreater{\N}{65}} },
            ]
            {dat_folder/err_eta.dat}{}
            {\\ \cline{2-4}
            & \N & \etad & \etao}
            \\
            \hline
        \end{tabular}
    \caption{$\|\eta-\eta^*\|_\infty/\|\eta^*\|_\infty$.}
    \label{tab:err_eta}
    \end{subtable}
    \hspace{1em}
    \begin{subtable}[h]{0.45\textwidth}
        \centering
        \begin{tabular}{|c|c|m{5em}|m{5em}|}%
            \hline
            \bfseries \% noise & \bfseries N & \bfseries Direct & \bfseries Optim.\\
            \hline
            \begin{filecontents*}{dat_folder/time.dat}
perErr,N,to,td
0,64,3.8907e-02,3.6850e-04
0,128,1.2486e-01,3.0548e-03
0,256,8.9789e-01,1.2311e-02
0,512,4.9093e+00,5.7834e-02
1,64,3.6985e-01,2.1128e-03
1,128,1.7341e+00,1.1814e-03
1,256,4.2470e+00,3.9572e-03
5,64,4.8052e-01,1.0649e-01
5,128,2.0892e+00,1.1955e-01
5,256,7.3085e+00,1.2354e-02
10,64,6.1017e-01,7.1150e-03
10,128,2.5986e+00,2.6298e-02
10,256,1.3278e+01,2.7406e-01
\end{filecontents*}
\csvreader[
            head to column names, 
            filter expr = { test{\ifnumequal{\perErr}{0}}
                            and test{\ifnumequal{\N}{64}} },
            ]
            {dat_folder/time.dat}{}
            {\multirow{4}{*}{\perErr} & \N & \td & \to }
            \csvreader[
            head to column names, 
            filter expr = { test{\ifnumequal{\perErr}{0}}
                            and test{\ifnumgreater{\N}{65}} },
            ]
            {dat_folder/time.dat}{}
            {\\ \cline{2-4}
            & \N & \td & \to}
            \\
            
            \hline
            \csvreader[
            head to column names, 
            filter expr = { test{\ifnumequal{\perErr}{1}}
                            and test{\ifnumequal{\N}{64}} },
            ]
            {dat_folder/time.dat}{}
            {\multirow{3}{*}{\perErr} & \N & \td & \to }
            \csvreader[
            head to column names, 
            filter expr = { test{\ifnumequal{\perErr}{1}}
                            and test{\ifnumgreater{\N}{65}} },
            ]
            {dat_folder/time.dat}{}
            {\\ \cline{2-4}
            & \N & \td & \to}
            \\
            
            \hline
            \csvreader[
            head to column names, 
            filter expr = { test{\ifnumequal{\perErr}{5}}
                            and test{\ifnumequal{\N}{64}} },
            ]
            {dat_folder/time.dat}{}
            {\multirow{3}{*}{\perErr} & \N & \td & \to }
            \csvreader[
            head to column names, 
            filter expr = { test{\ifnumequal{\perErr}{5}}
                            and test{\ifnumgreater{\N}{65}} },
            ]
            {dat_folder/time.dat}{}
            {\\ \cline{2-4}
            & \N & \td & \to}
            \\
            
            \hline
            \csvreader[
            head to column names, 
            filter expr = { test{\ifnumequal{\perErr}{10}}
                            and test{\ifnumequal{\N}{64}} },
            ]
            {dat_folder/time.dat}{}
            {\multirow{3}{*}{\perErr} & \N & \td & \to }
            \csvreader[
            head to column names, 
            filter expr = { test{\ifnumequal{\perErr}{10}}
                            and test{\ifnumgreater{\N}{65}} },
            ]
            {dat_folder/time.dat}{}
            {\\ \cline{2-4}
            & \N & \td & \to}
            \\
            \hline
        \end{tabular}
    \caption{Computation time (in hours).}
    \label{tab:time_comput}
    \end{subtable}
    \caption{Relative errors on the hydrodynamical variables and computation time, with or without noise, at different values of $N$ and for the direct and optimisation recovery procedures.}
    \label{tab:errors}
\end{table}